# Study of diamagnetism in laser-produced plasma using B-dot probe

Narayan Behera*, R. K. Singh, G. Veda Prakash, Kiran Patel, H. C. Joshi and Ajai Kumar

Institute for Plasma Research, Gandhinagar - 382 428, India

*Email: nbehera@ipr.res.in

## Abstract

Time-varying diamagnetism in laser-produced plasma moving across the transverse magnetic field with different field strengths has been studied using fast imaging and magnetic probe. The emphasis of the present work is on the development of suitable B-dot probe and quantitative analysis of induced diamagnetic field in an expanding plasma plume. A Helmholtz coil associated with pulsed power system is used to produce uniform magnetic field of different strengths. Helmholtz coil allowed the plume imaging along the magnetic field lines, which gives the direct structural information of the induced diamagnetic cavity. A high frequency three-axis B-dot probe has been developed to measure the transient magnetic field. Different experimental approaches have been used to test the response, sensitivity and calibration of the developed probe. Findings indicate that plasma plume shows the perfect diamagnetic behaviour for the external field varying from 0.13 T to 0.57 T where the induced magnetic field almost completely displaced the external magnetic field.

# 1 Introduction

The interaction of laser-produced plasma with external magnetic field is an important area because of its applications in many fundamental as well as applied researches, such as manipulation of plasma plume characteristics[1-3], increase in the detection sensitivity of laser-induced breakdown spectroscopy [4,5], debris mitigation [6] and laboratory investigation of astrophysical plasma. Several works have been carried out related to plasma plume expansion in uniform [7-11] and non-uniform magnetic fields [12] to understand the associated phenomena in plasma plume-magnetic field interactions, like plume confinement, enhancement of plume emissivity, plasma oscillations [13], edge instability and dramatic structuring [14-17] and sub-Alfvenic plasma expansion [18].

Diamagnetism in plasma plume is one of the important aspect in plasma plume expansion across the transverse magnetic field. When the laser-produced plasma expands in the presence of external magnetic field, it produces a diamagnetic current on the surface of the plasma plume and this diamagnetic current generates an induced magnetic field opposite to the applied magnetic field direction. So the applied magnetic field is displaced by this induced magnetic field [15]. Several experimental investigations related to parametric study of diamagnetism of laser-produced plasma and its effect on the dynamics and structure of the plasma plume have been done with permanent bar magnets [7,9]. Fast imaging [7,9] and magnetic probe [19,20] were used as main diagnostics tool for these studies. Recently, with the introduction of Helmholtz coil and simultaneous two-directional imaging, formation, evolution and three-dimensional structure of diamagnetic cavity in laser-produced plasma are experimentally demonstrated [8]. However quantitative parametric study of the diamagnetic cavity, e.g. diamagnetic field in plasma plume is scarce in the literature which is essential to understand plasma plume-magnetic field interactions. B-dot probe is commonly used to measure the time-varying magnetic field. Since the laser plasma plume is highly transient and therefore the design and calibration of the magnetic probe with the desired sensitivity and frequency range is a challenging task.

Based on the above considerations, we have designed a high frequency, three-axis B-dot probe having bandwidth up to 10 MHz to measure the diamagnetic field in expanding plasma plume across the transverse magnetic field. Plasma plume expands in uniform magnetic field produced by the Helmholtz coil and has been imaged along the magnetic field lines. The magnetic probe is placed at the diamagnetic regime, decided by the time resolved images of the diamagnetic cavity, to measure the net magnetic field. The details of designing aspects and experimental technique to insure the response and calibration of the magnetic probe are briefly described. The observed magnetic field in diamagnetic region of the plasma plume is also discussed.

## 2 Design of the probe

### 2.1 Construction of probe

In order to design a suitable magnetic probe to measure the induced magnetic field in laser- produced plasma, following points has been considered. The response of the probe should be faster than the time scale of induced magnetic field in the plasma plume, the size of the probe should be in the scale of the ion Larmor radius and it should be sensitive to measure the desired magnetic fields etc. Based on the above considerations we have designed and fabricated the probe as follows.

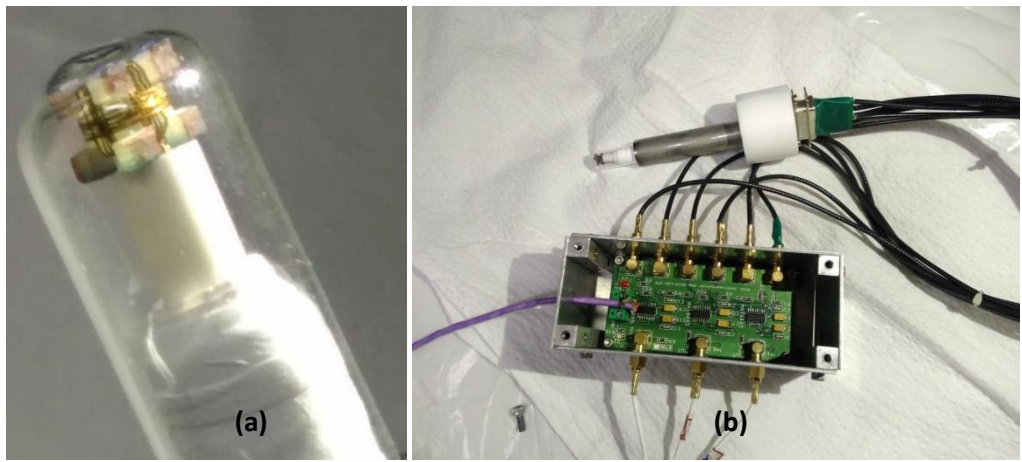

**Fig. 1.** (a) Probe head in a glass tube and (b) indigenously designed differential amplifier.

A 2.8 × 2.8 × 2.8 mm cubical G10 material is used to make a core for three-dimensional probes as shown in Fig. 1a. The G10 material is used due to its high thermal capacities, easy to machine and unity relative magnetic permeability at high frequencies. Five turns of a twisted pair of 40 gauge polyimide insulated copper wire is winded around the core which creates a differential pair along the each axis which helps to cancel the capacitive or electrostatic pickup in the signals. The probe is mounted on a long ceramic tube and the whole probe assembly is kept inside a glass test tube to protect it from the direct exposer of charged particles. Wires of the loops are connected with 50 Ω coaxial cables through 15 pin D-sub connectors and fed to an indigenously designed high frequency differential amplifier through SMA connectors (see, Fig. 1b). The details of high frequency differential amplifier (MAX4445, 550MHz-3dB Bandwidth) for eliminating the electrostatic/capacitive pickup is shown in Fig. 2a. The frequency response of the amplifier is verified by function generator (AFG 3102, Tektronix, 100 MHz, 1 GS/s sampling rate) and is shown in Fig. 2b. The output of differential amplifier is recorded on an oscilloscope with a termination impedance of 50 Ω.

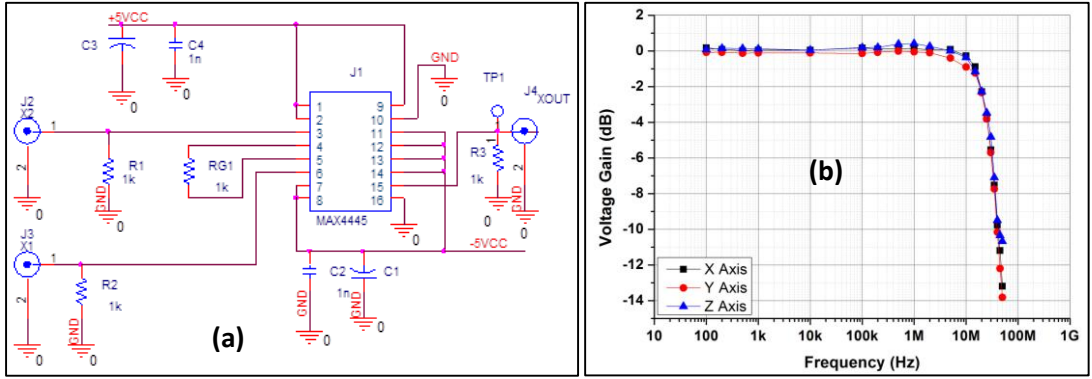

**Fig. 2** (a) Circuit schematics of a differential amplifier for one channel and (b) the voltage gain vs. frequency response of the amplifier.

The time-varying-magnetic field produced by the laser-produced plasma can be measured with the B-dot probe using the relation,

$$B = - \frac{1}{2gN_{Probe}A_{Probe}} \int V_{meas}\, dt \qquad (1)$$

where, $N_{probe}$ and $A_{probe}$ are the number of turns and cross-sectional area of the probe, $V_{meas}$ is the induced voltage and $g$ is the gain of amplifier. The factor $(2g\, N_{probe}\, A_{probe})^{-1}$ is the calibration factor for B-dot probe which can be obtained by measuring the response of the B-dot probe in known time-varying magnetic field. The magnitude of magnetic field is estimated by integrating the induced voltage signal either by using an electronic integrator or a numerical integrator. Here it should be noted that, in the present experiment, the frequency of the applied pulsed magnetic field is in the range of KHz whereas the magnetic field produced by the plasma plume lying the MHz range. In order to estimate both applied and plasma magnetic fields simultaneously, the numerical integrator is more appropriate choice because electronic integrator with such a high dynamic range (KHz to MHz) is not feasible.

## 2.2 Response of the probe

It is essential to measure the response of the probe to check if the probe is functioning correctly in the desired frequency range. As the phenomena like demagnetization in laser-produced plasma occurs at time scale of 200-1000 ns from the plasma initiation, the designed probe should respond to a few MHz range. We have adopted two approaches to test the response of the B-dot probe. In the first approach, response of probe is tested using pulsed magnetic field produced by Helmholtz coil, where the linearity of the probe voltage is tested with respect to frequency of the Helmholtz coil current, keeping the current constant. In the second approach, the response of the probe is verified by the magnetic field produced by the circular inductor coil where the phase matching between the inductor current and induced magnetic field is monitored. The details of the two approaches are described in the following sections.

**2.2.1 With Helmholtz coil**

The response of the probe is tested with Helmholtz coil configuration in frequency range 50 Hz-10 MHz using the scheme shown in Fig. 3.

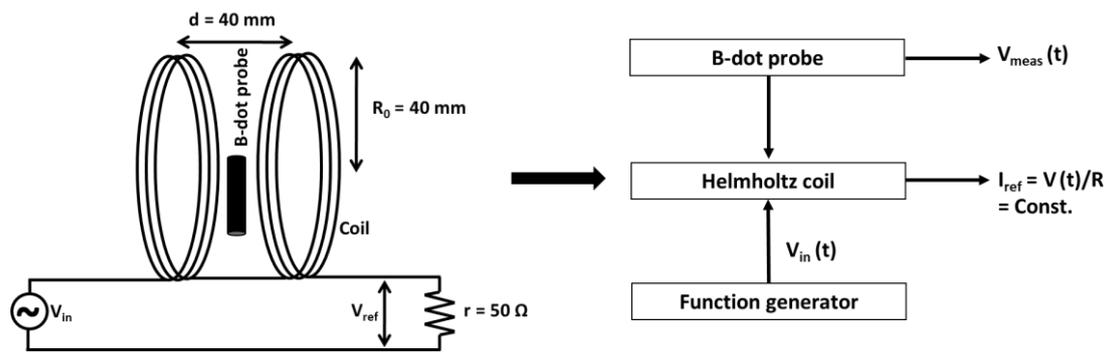

**Fig. 3.** Scheme to check the response of B-dot probe using Helmholtz coil.

The Helmholtz coil consists of two identical coils of 20 turns each and 40 mm radii which are separated by a distance of 40 mm. A time-varying sinusoidal voltage of frequencies 500 Hz-1 MHz and constant peak to peak amplitude $V_{ref}$ = 1.52 V is applied to the Helmholtz coil by a function generator. A resistance R = 50 Ω is connected in series with the coils. The magnetic field produced at the centre of the coil is $B = \frac{8}{5\sqrt{5}} \frac{\mu_0 n I}{R_0}$. In this case, the current can be estimated as voltage drop across the resistor divided by the 50 Ω. In this configuration of the Helmholtz coil, for 1 A current flow, the magnetic field produced by the coil at the centre is 4.5 G. The B-dot probe is kept at the centre of the Helmholtz coil. The output voltage $V_{meas,pp}$ at different input frequencies are measured using a Digital Oscilloscope and are plotted as a function of frequencies from 500 Hz to 1 MHz as shown in Fig. 4a. The linear variation of $V_{meas,pp}$ with frequencies shows the proper response of probe in 500 Hz-1 MHz. At higher frequencies (1-10 MHz range) the input peak to peak voltage $V_{ref}$ is reduced to 840 mV, and corresponding $V_{meas,pp}$ vs frequency plot as shown in Fig. 4b which shows the linear response of the designed B-dot probe in 1-10 MHz. However as we are using low input voltages which may introduce uncertainty in $V_{meas}$. Therefore in MHz range response of the probe is reinsured with circular inductor which will be discussed in next section.

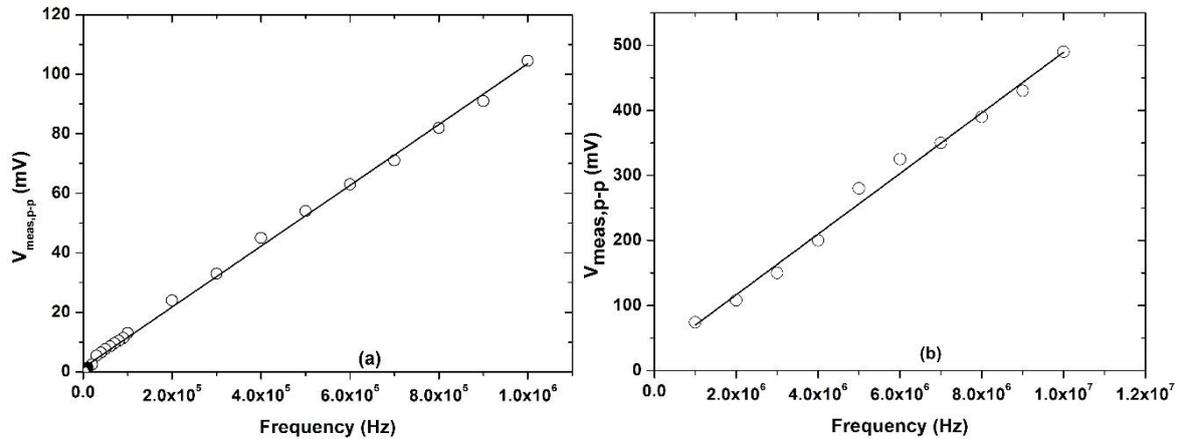

**Fig. 4.** The frequency response from (a) 500 Hz to 1 MHZ and (b) 1 MHz to 10 MHZ by the magnetic field produced by the Helmholtz coil.

**2.2.2 With circular inductor**

As already discussed, at higher frequency, it is difficult to maintain the desired current level in the Helmholtz coil to cheek the response of the probe with acceptable uncertainty. Circular inductor is better choice to check the response of the probe in the MHz frequency range.

The probe is said to be responding perfectly at a given frequency if the inductor current and induced magnetic field waveforms are in the same phase and hence 90° phase difference with input current profile and induced B-dot probe voltage profile. This basic principle is used to verify the response of the probe at higher frequencies with improved sensitivity (Fig. 5). The experimental scheme for that has been summarized in Fig. 5.

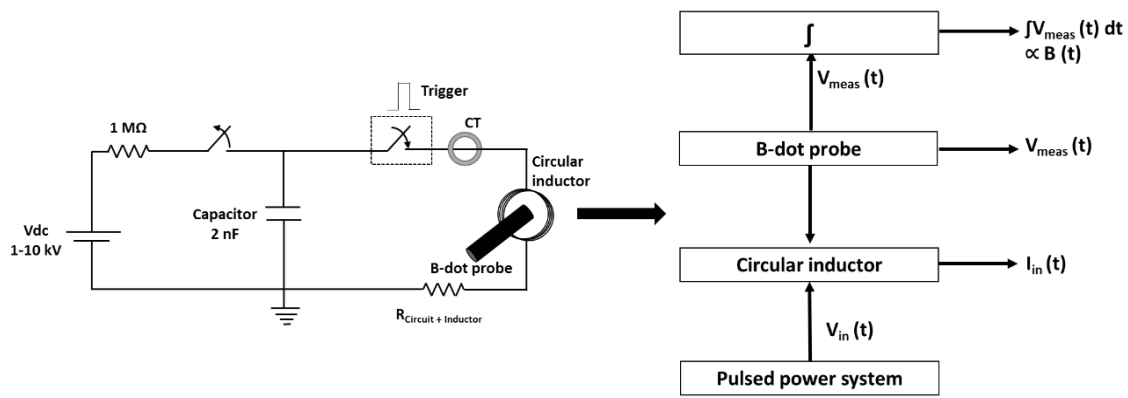

**Fig. 5.** Scheme and setup for checking the probe response using the circular inductor.

A circular inductor is made of 0.8 mm wire wound 10 turns in a circle of radius 22 mm. A pulses power system which consist of DC power supply and 2 nF capacitor bank has been used as a pulsed current source for the circular inductor as shown in Fig. 5. For a preset charging voltage of 150 V, the discharge of the capacitor produces 2 A current in a circular inductor and consequently generate the magnetic field ~ 7 G. The B-dot probe is placed at the centre of the coil. A current transformer (CT) is used to measure the current in the circular inductor. Current profiles of both the inductor and probe signal (induced voltage profile) are monitored on an oscilloscope.

Typical time-varying current and induced voltage profiles of the B-dot probe as a function of time are shown in Fig. 6. Numerical integration of the induced voltage using MATLAB code which represents the induced magnetic field profile is also included in Fig. 6. Fig. 6 clearly shows that induced magnetic field waveform is in same phase with input current wave form and also there is 90º phase difference between induced voltage waveform and input current profile. This observation confirms that the designed probe perfectly responds for the magnetic flux at 1.27 MHz frequency.

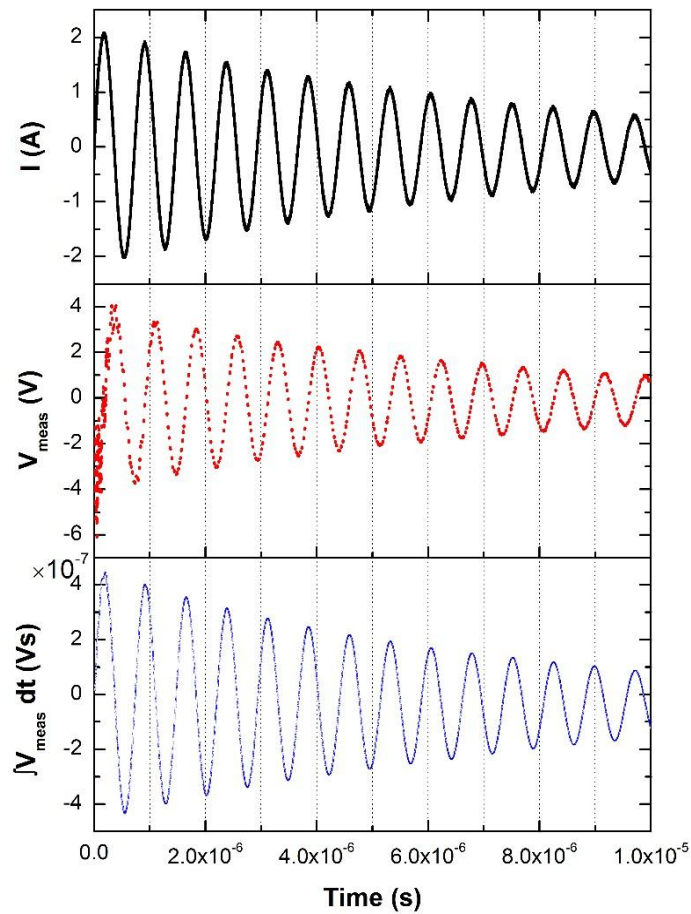

**Fig. 6.** The phase matching of current, $V_{meas}$ and $\int V_{meas}\,dt$ at 2.0 A and 1.27 MHz.

In summary, the developed B-dot probe responds perfectly in desired frequency range (500 Hz-10 MHz) as verified by both Helmholtz coil and circular inductor.

## 2.3 Calibration of the probe

After ensuring the response of the magnetic probe at desired frequency range, it is necessary to calibrate the probe with known magnetic field for absolute measurement of the magnetic field. As we have already mentioned that in MHz frequency range, Helmholtz coil is not suitable for calibration. Therefore we have used a pulsed power supply coupled circular inductor as described earlier to calibrate the B-dot probe at 1.27 MHz.

The magnetic field at the centre of the circular inductor is given by

$$B\ (x = 0) = \frac{1}{2}\mu_0 nI \left( \frac{x-a_1}{\sqrt{(x-a_1)^2 + R^2}} - \frac{x-a_2}{\sqrt{(x-a_2)^2 + R^2}} \right) \quad (2)$$

Here x is the position along the axis of inductor where magnetic field is calculated (e.g., x = 0 for magnetic field at centre), $\mu_0 = 4\pi \times 10^{-7}$ H/m is the vacuum permeability, n is turns per unit length, I is the electric current, R is radius of the inductor and $a_1$ and $a_2$ are lengths of left and right ends of the inductor from x respectively (here $a_1 = - L/2$ and $a_2 = L/2$, where L is the length of inductor). For different charging voltages of the capacitor (or different input currents in the inductor) the magnetic field strength at the centre of the inductor is estimated using relation (2). The B-dot probe is used to measure the induced voltage $V_{meas}$ for different magnetic field strengths. The integrated induced voltage that is $\int V_{meas}(t)\ dt$ is plotted as function of test magnetic field as shown in Fig. 7.

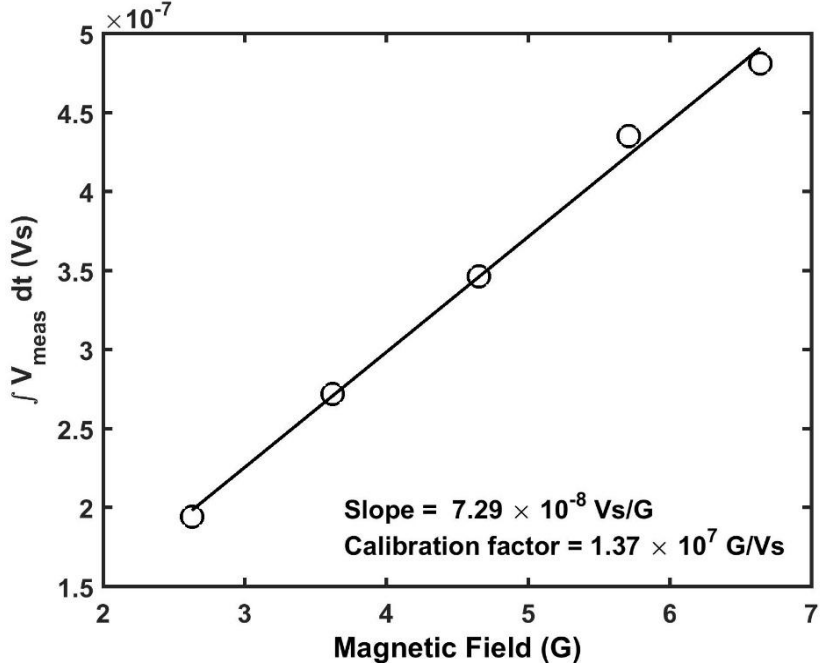

**Fig. 7.** Variation of ∫V$_{meas}$(t) dt as a function applied magnetic field. Reciprocal of the slope gives the calibration factor of the B-dot probe.

The calibration factor of the probe is obtained by the inverse of slope of the linear fit of the experimental data. The calibration factor is found to be $1.37 \times 10^7$ G/Vs which is used to estimate the unknown magnetic field. The estimated statistical error in the measured magnetic field is within 10 %.

## 3 Experimental scheme

Detailed description of the experimental setup has been presented in previous report [8]. Only a brief summary, which is important for the present study, is presented here. The schematic of the experimental set-up is shown in Fig. 8. The plasma plume is created in vacuum chamber having a base pressure less than $10^{-6}$ Torr using a 1064 nm, 8 ns pulsed Nd:YAG laser. The pulse energy and spot size of the laser beam are set to 150 mJ and 1 mm in diameter at the target, respectively, which corresponds to the laser fluence 19.1 J/cm$^2$ at the target surface. Transverse magnetic field (0.13-0.57 T) is produced by a Helmholtz coil with the indigenously developed pulse power supply. A polished aluminium sheet of thickness 2 mm is used as a target and it was placed in between the Helmholtz coils. An ICCD camera along the magnetic field is used to identify the position of the diamagnetic cavity as well as position of B-dot probe in the chamber. The designed B-dot probe is placed at the expected diamagnetic cavity (position is inferred from ICCD camera images), which can measure the net magnetic field ($\Delta B_{plasma} = B_{plasma} - B_{diamganetic}$) or magnetic field produced by plasma ($B_{plasma}$) in this region. Flexible bellow based linear feed through is used to vary/fix the position of B-dot probe. A high frequency Digital Oscilloscope is used to record time-varying induced voltage ($V_{meas}$).

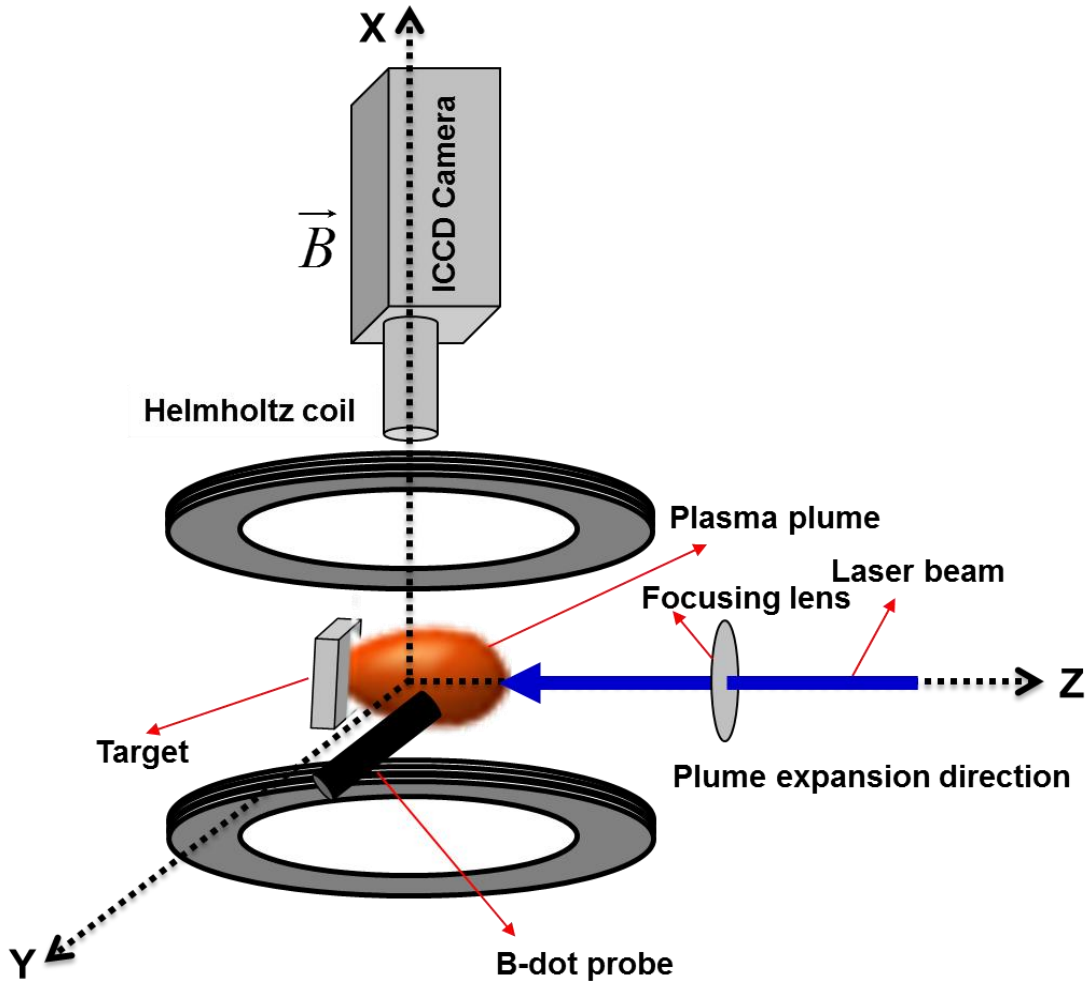

**Fig. 8.** Schematic of the experimental set-up.

## 4 Results and discussion

Diamagnetic cavitation in expanding plasma plume across the transverse magnetic field is one of the important phenomena in plume-magnetic field interaction. The characteristics of diamagnetism in plasma plume, i.e. formation, evolution and collapse of magnetic bubble depends on the plasma parameters (electron temperature and density) and applied magnetic field strength. To visualize the diamagnetism in expanding plasma plume across the magnetic field, emitting plume is imaged by the ICCD camera along the magnetic field lines. Two-dimensional images of the expanding plasma plume in the absence of magnetic field in vacuum and at time delay 200, 400 and 800 ns from the plasma initiation are shown in Fig. 9a. The laser fluence is set as 19.1 J/cm$^2$. This image represents the spectrally integrated emission intensity profile of emitting plume species in the range of 350-800 nm. Conventional adiabatic expansion of ellipsoidal shape plume [21] is clearly observed in the absence of the external magnetic field as shown in Fig. 9a where the electron temperature and density and hence emission intensity of plume are significantly decrease at higher time delays. On the other hand, shape, size and intensity of the plasma plume are completely modified on introduction of 0.13 T transverse magnetic field as shown in Fig. 9b. Fig. 9b represents two-dimensional images of plume along the plane perpendicular to the magnetic field lines at 0.13 T magnetic field, all other experimental parameters are same as in the case of absence of magnetic field. Initially, the cavity expands with time and a well-shaped elliptical cavity (magnetic bubble) is observed at the time delay 200 ns as shown in Fig. 9b. The cavity attains maximum dimension at time delay 400 ns. At delay time greater than 400 ns, plasma cavity starts shrinking mainly along the lateral direction, and finally, the cavity-like structure disappears with a further increase in time delay. A typical collapsing cavity is shown at a time delay of 800 ns in Fig. 9b.

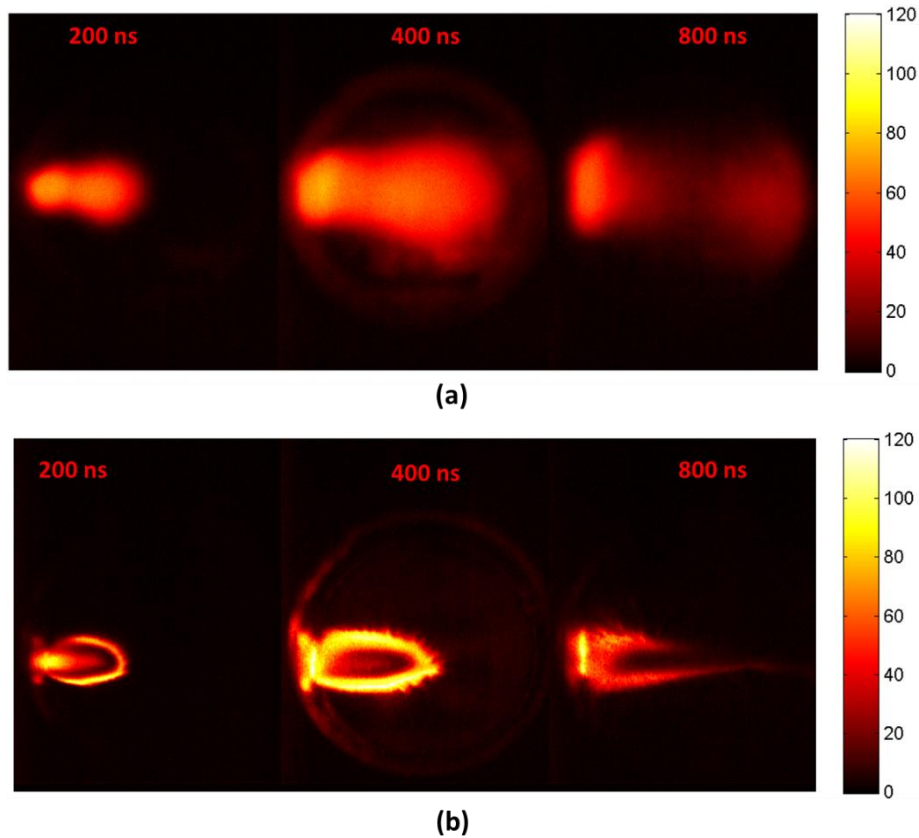

**Fig. 9.** Sequence of images of the expanding plasma plume (a) without magnetic field and (b) in the presence of 0.13 T magnetic field at 200, 400 and 800 ns delay times.

The diamagnetism of plasma plume could be understood as follows [15]. Ions and electrons in the plasma plume experience Lorentz force in the presence of magnetic field and therefore gyrate in opposite direction with different radii. Generally Larmor radius for ions is greater than the plume dimension and hence expands radially outward. On the other hand magnetized electrons tie to the magnetic field and try to hold the ions back and eventually produce radially inward electric field ***E***. As a result plasma get compressed into a thin shell and from a magnetic bubble [15]. Due to influence of ***E***×***B*** force, electrons are drifted azimuthally and generate the diamagnetic current. This current generates the induced magnetic field opposite to the applied magnetic field. The induced magnetic field displaced the applied magnetic field efficiently until the magnetic bubble is collapsed.

The diamagnetic nature of the plasma plume means the magnitude of induced magnetic field is comparable to applied magnetic field. To study the diamagnetism of the plasma plume, we have measured the induced magnetic field by B-dot probe. Based on the recorded plume images, B-dot probe is placed at the diamagnetic region. The typical three-axis magnetic probe signal in absence external magnetic field is shown in Fig. 10. $V_{x\text{-probe}}$, $V_{y\text{-probe}}$ and $V_{z\text{-probe}}$ represent the signals produced by self-generated magnetic fields in the expanding plasma plume along the along x-axis, y-axis and z-axis respectively as described in the schematic of the setup (Fig. 8). Fig. 11 shows the probe signals along the x-axis in the presence of 0.13 T external magnetic field and in the absence of plasma plume. The induced voltage profile corresponding to the external magnetic field is clearly observed in $V_{x\text{-probe}}$ signal. The small fraction of signal detected by $V_{y\text{-probe}}$ and $V_{z\text{-probe}}$ may be due to slight variation in orthogonal placement of probe and hence causes the leakage of field component. $V_{x\text{-probe}}$ profile is used to synchronize the pulsed magnetic field with the plasma formation. The magnetic field is pre-triggered in such a way that zero crossing point (maxima of magnetic field) and laser pulse coincide at t = 0.

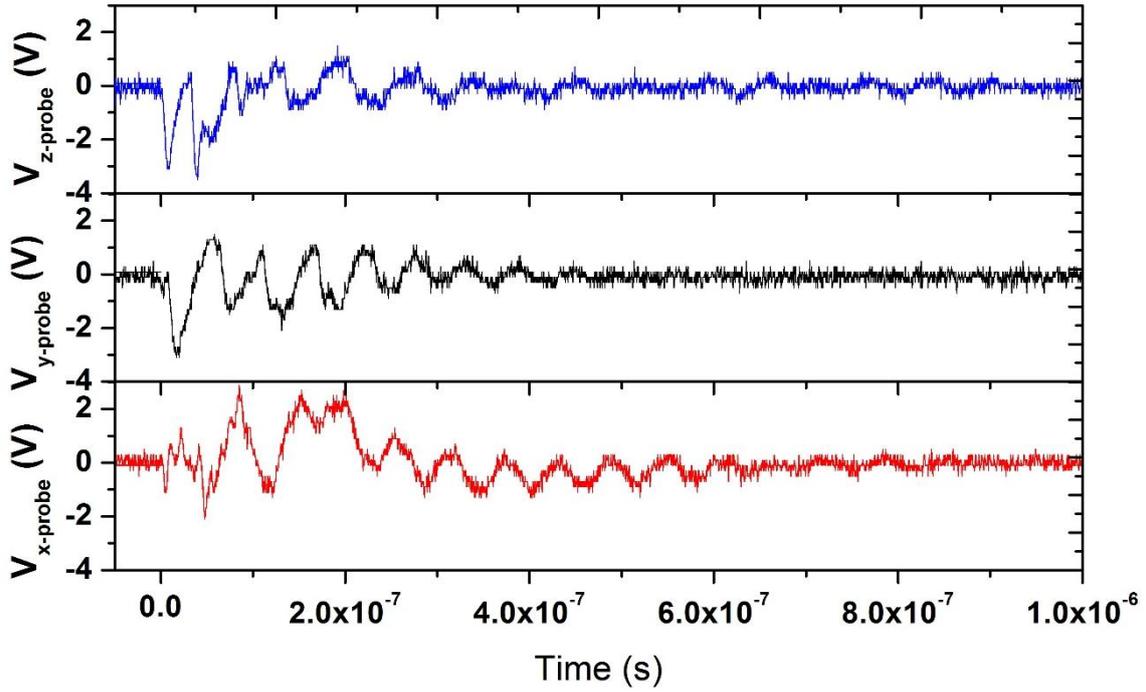

**Fig. 10.** Time profile of induced voltage in x, y and z axes of B-dot probe in the absence of external magnetic field.

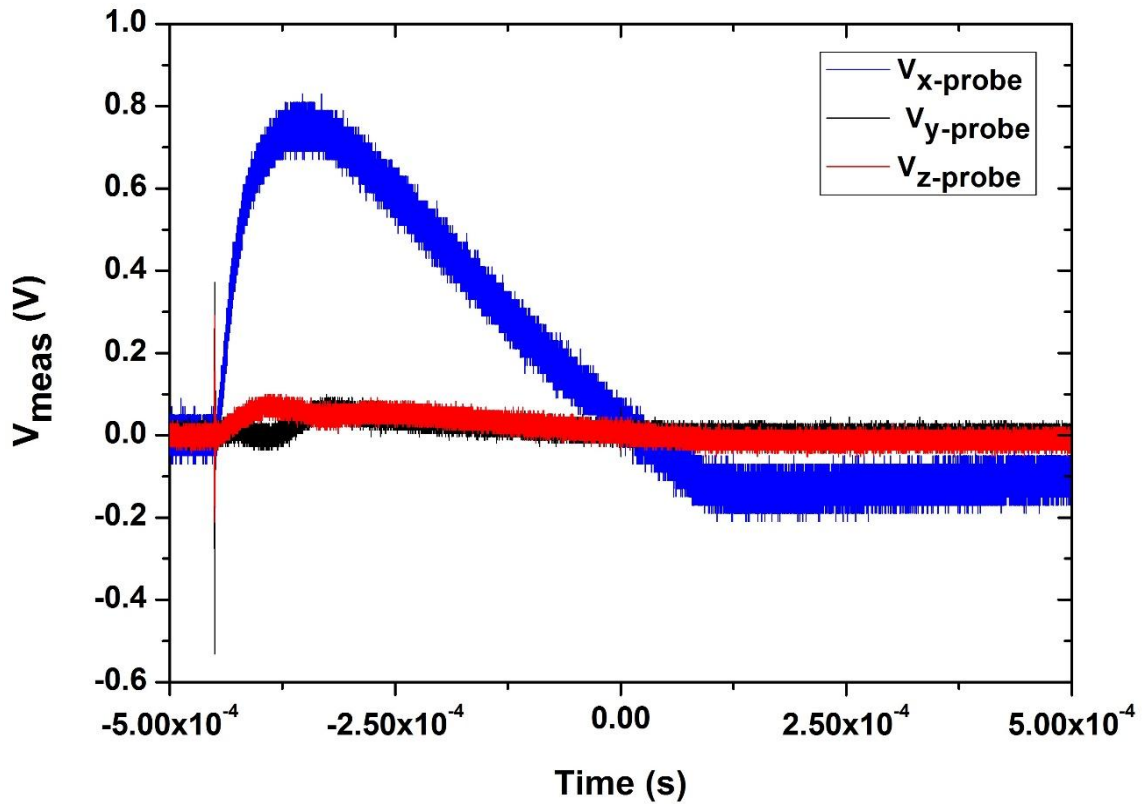

**Fig. 11.** Time profile of induced voltage of B-dot probe at 0.13 T applied external magnetic field in the absence of plasma plume.

The plasma plume and magnetic field interaction and subsequent diamagnetism of plume are clearly demonstrated in Fig. 12 where $V_{x-probe}$ signal plotted as a function of time in the presence of both plasma plume and 0.13 T external magnetic field. Apart from the slow induced voltage correspond to the applied magnetic field, an additional high frequency signal appeared at the onset of plasma formation. This high frequency signal arises due to the generation of diamagnetic current and hence the magnetic field under the influence of external magnetic field. For clarity, zoomed signal for induced magnetic field is shown in inset of Fig. 12. Here it should be noted that the induced diamagnetic field is opposed and displaced the applied magnetic field and therefore $V_{x-probe}$ measures the net magnetic $\Delta B_{x-plasma}$, that is the difference between external applied magnetic field and diamagnetic magnetic field produced by plasma plume. In a perfect diamagnetic behaviour of the plasma plume, magnitude of induced magnetic field is comparable to the applied magnetic field and the net magnetic field $\Delta B_{x-plasma}$ is tending to zero. By applying the integration and calibration procedure as described in previous section on the $V_{x-probe}$ signal (shown in inset of figure), the estimated net magnetic field $\Delta B_{x-plasma}$ is ~ 9 G. The negligible net magnetic field $\Delta B_{x-plasma}$ in comparison to 0.13 T external magnetic field indicates that at considered time delay plasma plume shows the diamagnetic behaviour where induced magnetic field almost completely displaced the external magnetic.

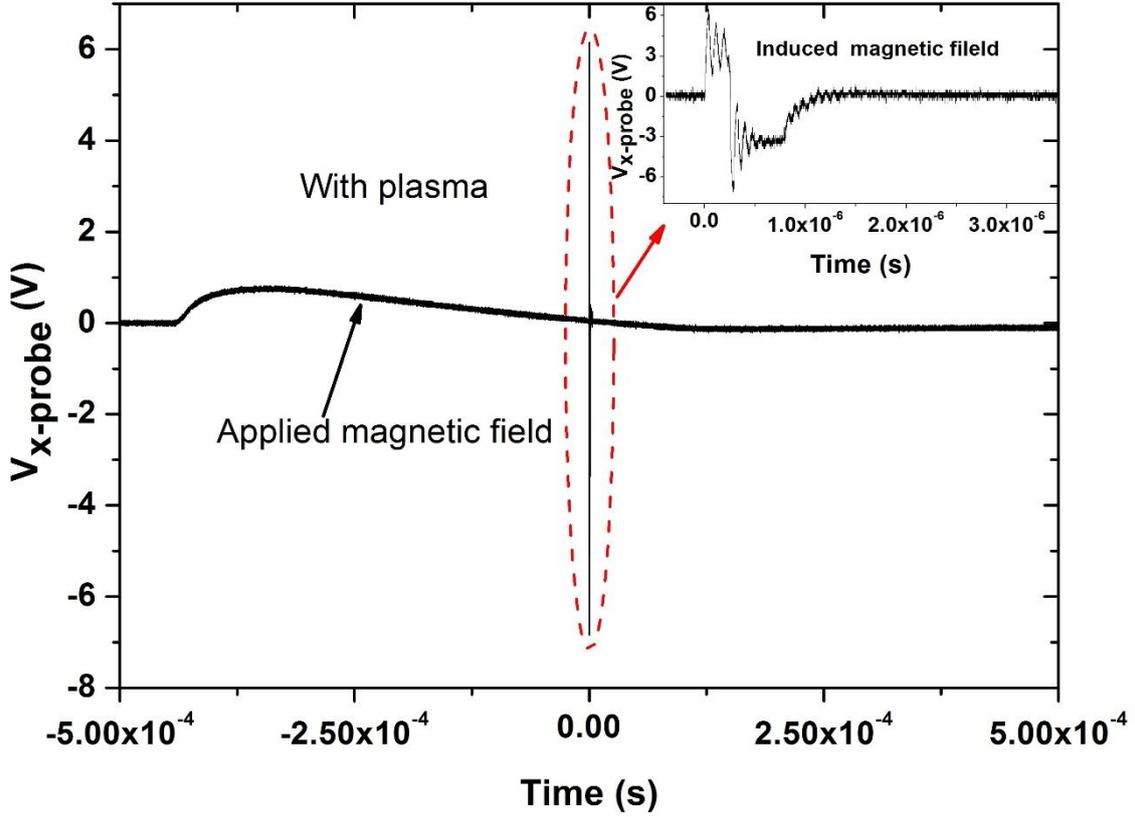

**Fig. 12.** Time profile of $V_{x\text{-probe}}$ signal in presence of plasma plume at 0.13 T magnetic field.

In order to further validate the diamagnetism in plasma plume, the induced magnetic field profile ($V_{x\text{-probe}}$ voltage profiles) are recorded with different external magnetic field intensities, varying from 0.13 T to 0.57 T as shown in Fig. 13a. These high frequency voltage signals are integrated using the numerical integrator to obtain corresponding net magnetic field profile ΔB along applied external magnetic field (x-axis) which is depicted in Fig. 13b.

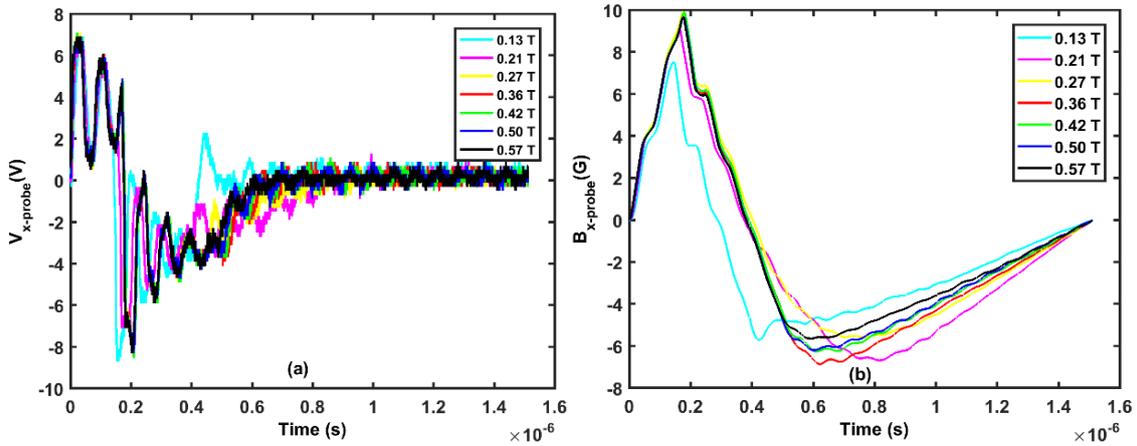

**Fig. 13.** Time profile of (a) Vx-probe and (b) magnetic field obtained by integrating voltage signals in the presence of magnetic field ranging from 0.13 T to 0.57 T.

For better clarity, the magnitude of net magnetic field is plotted as function of external magnetic field intensity as shown in Fig. 13. Fig. 14 clearly shows that the estimated net magnetic field are found to be in the range of 7-9 G which is around ~ 0.1% of the applied magnetic field ranging from 0.13 to 0.57 T. This simply means that for the present experimental conditions and for any applied magnetic field, the induced diamagnetic field is almost comparable to external magnetic field and hence completely displaced the applied magnetic field. The above observations confirm the prefect diamagnetic behaviour of the plasma plume in the considered magnetic field range.

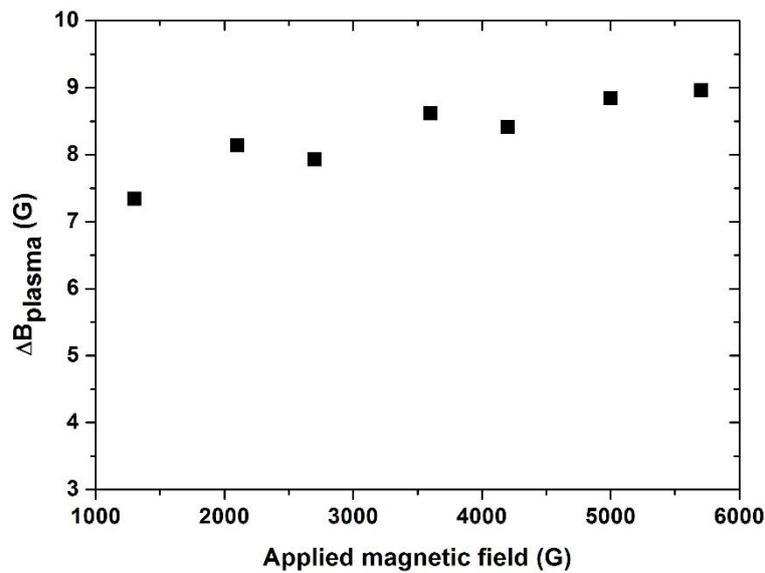

**Fig. 14.** The variation of induced net magnetic field measured by B-dot probe along the magnetic field lines (x-axis) with different applied external magnetic fields.

# 5 Conclusion

A high frequency, three-axis B-dot probe has been developed for the study of diamagnetism of laser-produced plasma expanding across the external transverse magnetic field. Technical aspects of design, fabrication, signal processing and acquisition are briefly discussed. The experimental procedure for the calibration and validation of response of the probe for a wide frequency range using the Helmholtz coil and circular inductor are also described in details. Laser-produced aluminium plasma plume expansion across the variable transverse magnetic field and subsequent formation of diamagnetic cavity is imaged along a plane perpendicular to the applied magnetic field lines using the ICCD camera. The ICCD images are clearly demonstrated the formation of diamagnetic cavity and its evolution in aluminium plasma plume. The B-dot probe is placed at the expected diamagnetic cavity region inferred from ICCD images to measure the resultant magnetic field $\Delta B_{plasma}$. The measured $\Delta B_{plasma}$ is found to be in the range 7-9 G for in the presence of external magnetic field ranging from 0.13 to 0.57 T indicates that induced magnetic field is almost completely displaced the external magnetic field and confirmed the diamagnetic behaviour of the plasma plume. This work may be useful in studying magnetohydrodynamics (MHD) in magnetized plasmas.

## Acknowledgement

The author NB would like to thank **Dr. Amulya Kumar Sanyasi** for his help in various aspects of B-dot probe.